\documentclass[conference]{IEEEtran}
\IEEEoverridecommandlockouts
\usepackage{cite}
\usepackage{amsmath,amssymb,amsfonts}
\usepackage{algorithmic}
\usepackage{graphicx}
\usepackage{textcomp}
\usepackage{subcaption}
\usepackage{xcolor}
\usepackage{multirow,tabularx}
\usepackage{booktabs}
\usepackage[normalem]{ulem}
\def\BibTeX{{\rm B\kern-.05em{\sc i\kern-.025em b}\kern-.08em
    T\kern-.1667em\lower.7ex\hbox{E}\kern-.125emX}}
\makeatletter

\begin{document}

\title{Learned Video Compression for YUV 4:2:0 Content Using Flow-based Conditional Inter-frame Coding}

\author{
    \begin{tabular}{cccc}
        Yung-Han Ho$^{1}$ &
        Chih-Hsuan Lin$^{1}$ &
        Peng-Yu Chen$^{1}$ & 
        Mu-Jung Chen$^{1}$\\
    \end{tabular}\\
    \begin{tabular}{ccc}
        Chih-Peng Chang$^{1}$ &
        Wen-Hsiao Peng$^{1}$ &
        Hsueh-Ming Hang$^{2}$\\
    \end{tabular}\\
    
    $^1$Computer Science Dept., 
    $^2$Electronics Engineering Dept., 
     National Yang Ming Chiao Tung University, Taiwan
    \vspace{-2ex}
}


\maketitle

\begin{abstract}
This paper proposes a learning-based video compression framework for variable-rate coding on YUV 4:2:0 content. Most existing learning-based video compression models adopt the traditional hybrid-based coding architecture, which involves temporal prediction followed by residual coding. However, recent studies have shown that residual coding is sub-optimal from the information-theoretic perspective. In addition, most existing models are optimized with respect to RGB content. Furthermore, they require separate models for variable-rate coding. To address these issues, this work presents an attempt to incorporate the conditional inter-frame coding for YUV 4:2:0 content. We introduce a conditional flow-based inter-frame coder to improve the inter-frame coding efficiency. To adapt our codec to YUV 4:2:0 content, we adopt a simple strategy of using space-to-depth and depth-to-space conversions. Lastly, we employ a rate-adaption net to achieve variable-rate coding without training multiple models. Experimental results show that our model performs better than x265 on UVG and MCL-JCV datasets in terms of PSNR-YUV. 
However, on the more challenging datasets from ISCAS'22 GC, there is still ample room for improvement. This insufficient performance is due to the lack of inter-frame coding capability at a large GOP size and can be mitigated by increasing the model capacity and applying an error propagation-aware training strategy.



\end{abstract}


\section{Introduction}

Since deep neural networks have demonstrated their great potential in computer vision tasks, learning-based video compression has rapidly risen in recent years. DVC~\cite{dvclu} is the first work that integrates neural networks with the predictive coding concepts for video compression. Following works like M-LVC~\cite{mlvc} and HLVC~\cite{hlvc} utilize multi-reference frames to improve the coding efficiency. Furthermore, FVC~\cite{fvc} performs predictive coding operations in the feature domain with the deformable convolution. ELV-VC~\cite{elfvc} proposes to effectively send the incremental flow based on the flow map predictor. Nevertheless, several issues remain unsolved for learning-based video compression.

First of all, the effectiveness of the residual coding is a concern, and the learning-based approach should provide more flexibility than traditional predictive coding. Ladune~\textit{et al.}\cite{mmsp} first point out the inefficiency of the residual coding from the perspective of  information theory. They explain that given the motion-compensated frame $x_c$ for coding the target frame $x_t$, the expected entropy of residual coding should be greater than or equal to the conditional coding
$H(x_t - x_c) \geq H(x_t - x_c | x_c) = H(x_t | x_c)$. To this end, they propose to use conditional VAE that concatenates the motion-compensated frame with the target frame and the latent features in the encoding and decoding processes. DCVC~\cite{dcvc} improves Ladune's work by replacing the motion-compensated frame with its latent representation. Additionally, a conditional temporal prior is introduced for better entropy coding. However, how to effectively use conditional information is still an issue to be discussed.  

Secondly, the use of a single model to implement variable-rate coding and rate control is also a challenge for learning-based video compression. Most learned video compression methods can only be optimized for a single rate point and will cause high memory consumption. Choi~\textit{et al.}\cite{cconv} propose a multi-rate image compression network with conditional convolution. Conditional convolution performs channel-wise scaling and shifting of the intermediate features. By replacing each convolutional layer with conditional convolution (CConv), it reprograms the feature to adapt to different dynamic ranges. For video compression, Lin~\textit{et al.}\cite{icip21} further apply the similar technique but without the shifting operation to both motion and residual coder. Though these research provide solutions for variable-rate image and video compression, it is still unable to achieve precise rate control.


Finally, most learning-based compression models operate on RGB color space, and YUV color format is more popular among actual video standards. To obtain better coding efficiency, how to deal with YUV 4:2:0 input format for learning-based video compression is still an open question.

Considering all the above issues, we propose a conditional flow-based video compression framework that uses YUV 4:2:0 video as input format. Our framework can also use only one model to adapt to multiple bit rates, and it can be extended to achieve rate control. The experimental results show that our method performs better than x265 on UVG~\cite{uvg} and MCL-JCV~\cite{mcl} datasets in terms of PSNR-YUV. 
However, on the more challenging datasets from ISCAS'22 GC, there is still ample room for improvement. We believe that this inferior performance is due to insufficient inter-frame coding at a large GOP size, which can be improved by increasing the model capacity and applying an error propagation-aware training strategy.

\section{Related Work: Augmented Normalizing Flow-based Image Compression (ANFIC)}

\begin{figure}
\centering
\begin{subfigure}{0.4\linewidth}
    \centering
    \includegraphics[width=\linewidth]{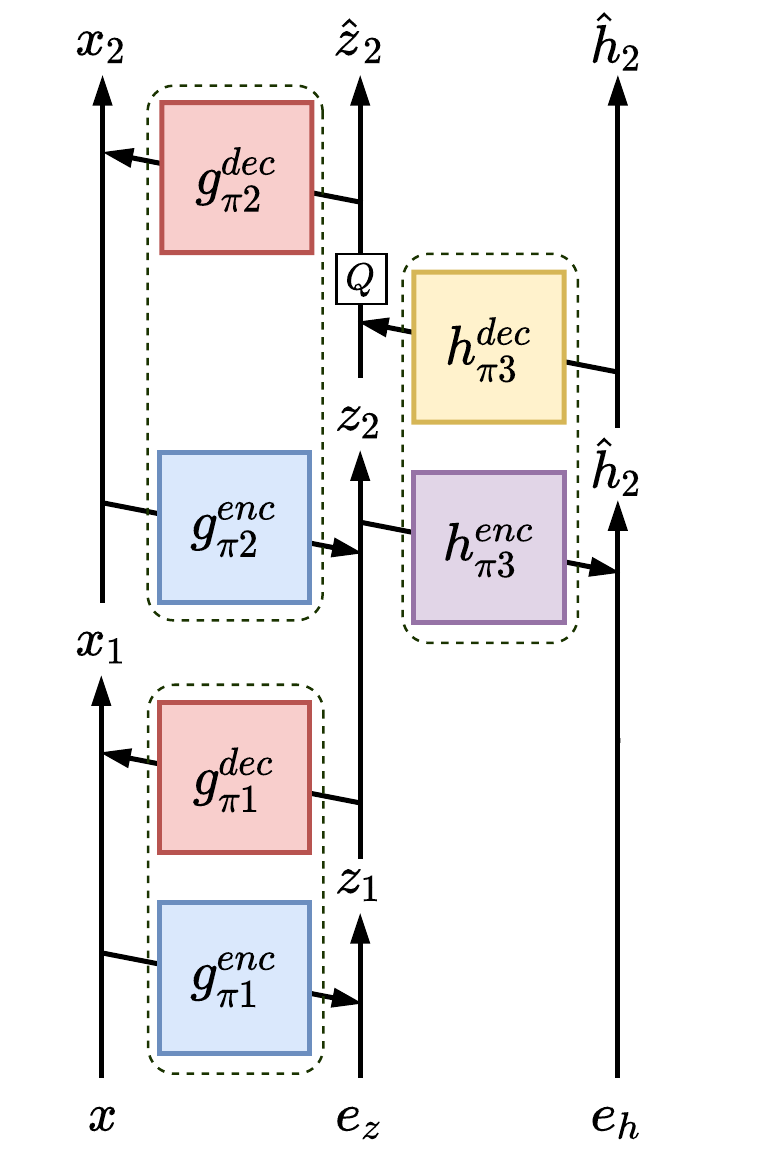}
    \vspace{-1.7em}
    \caption{}
    \vspace{-0.4em}
    \label{fig:anfic_a}
\end{subfigure}
\begin{subfigure}{0.45\linewidth}
    \centering
    \includegraphics[width=\linewidth]{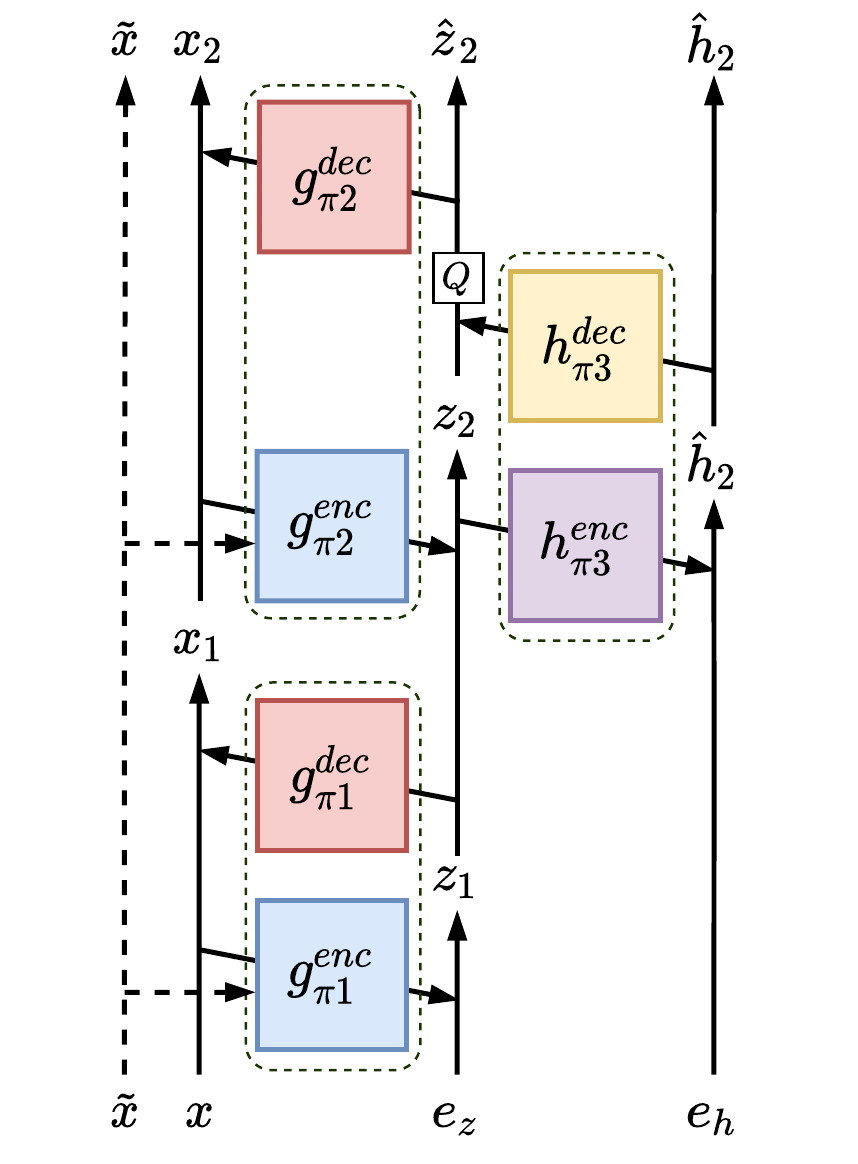}
    \vspace{-1.7em}
    \caption{}
    \vspace{-0.4em}
    \label{fig:anfic_b}
\end{subfigure}
\caption{The architectures of (a) ANFIC intra-frame coding~\cite{anfic} (N=128, M=192, K=320, L=192) and (b) the proposed conditional inter-frame coding (N=128, M=192, K=128, L=128), where N, M, K, L are the channel number of autoencoding transforms, hyperprior transform, $\hat{z}_2$, and $\hat{h}_2$.}
\label{fig:anfic}
\vspace{-0.3cm}
\end{figure}

ANFIC~\cite{anfic} is an image compression framework that leverages the VAE-based image compression in a flow-based model. Fig.~\ref{fig:anfic_a} illustrates the architecture of a 2-step ANFIC, which includes a stack of autoencoding transforms ($g_{\pi_1}^{enc},g_{\pi_1}^{dec}, g_{\pi_2}^{enc},g_{\pi_2}^{dec}$) and a hyperprior transform ($h_{\pi_3}^{enc},h_{\pi_3}^{dec}$).
ANFIC encodes an input image $x$ together with the augmented noise $e_z$, $e_h$ into the latent representation $(x_2, \hat z_2, \hat h_2)$. Taking $g_{\pi_1}^{enc},g_{\pi_1}^{dec}$ as an example, the transformation is defined as:
\begin{alignat}{2}
    g_{\pi_1}^{enc}(x, e_z) &= (x, e_z + m_{\pi_1}^{enc}(x))     &= (x,   z_1) \\
    g_{\pi_1}^{dec}(x, z_1) &= (x - \mu_{\pi_1}^{dec}(z_1), z_1) &= (x_1, z_1)
\end{alignat}
The second autoencoding transform $g_{\pi_2}^{enc},g_{\pi_2}^{dec}$ follows the same operations but takes $(x_1, z_1)$ as input.

As for the autoencoding transform of the hyperprior $h_{\pi_3}^{enc}, h_{\pi_3}^{dec}$, it follows~\cite{googleiclr18} for the entropy coding and the transformation can be written as: 
\begin{alignat}{2}
    h_{\pi_3}^{enc}(z_2, e_h) &= (z_2, e_h + m_{\pi_3}^{enc}(z_2)) &= (z_2, \hat h_2) \\
    h_{\pi_3}^{dec}(z_2, \hat h_2) &= (\lfloor z_2 - \mu_{\pi_3}^{dec}(\hat h_2) \rceil, \hat h_2) &= (\hat z_2, \hat h_2) 
\end{alignat}
where $\lfloor \cdot \rceil$ denotes the nearest-integer rounding operation (sketched as \fbox{$Q$} in Fig.\ref{fig:anfic}), and $m_{\pi}^{enc},\mu_{\pi}^{dec}$ are element-wise additive transformation parameters learned by the neural networks.

In a nutshell, ANFIC vertically stacks multiple autoencoding transforms for greater model expressiveness and horizontally extends an additional autoencoding transform of hyperprior for entropy coding. The latent variables $\hat z_2$ and $\hat h_2$ are expected to capture most of the information about the input $x$ and force $x_2$ to approximate $0$. Therefore, we only need to transmit $\hat z_2, \hat h_2$ while $x_2$ is replaced with $0$ during decoding.



\section{Proposed Method}

In this section, we describe our video compression system in detail. First, we present an overview of the proposed system, followed by an introduction to our conditional inter-frame coding and how it addresses the coding of YUV 4:2:0 content. Second, we show how the proposed system is extended to a variable-rate system for supporting variable-rate encoding without having to train separate networks. Lastly, we give the training procedure.

\subsection{System Overview}
\label{Sec:ArchitectureOverview}
\begin{figure}
\vspace{-.6em}
\centering
\includegraphics[width=0.8\linewidth]{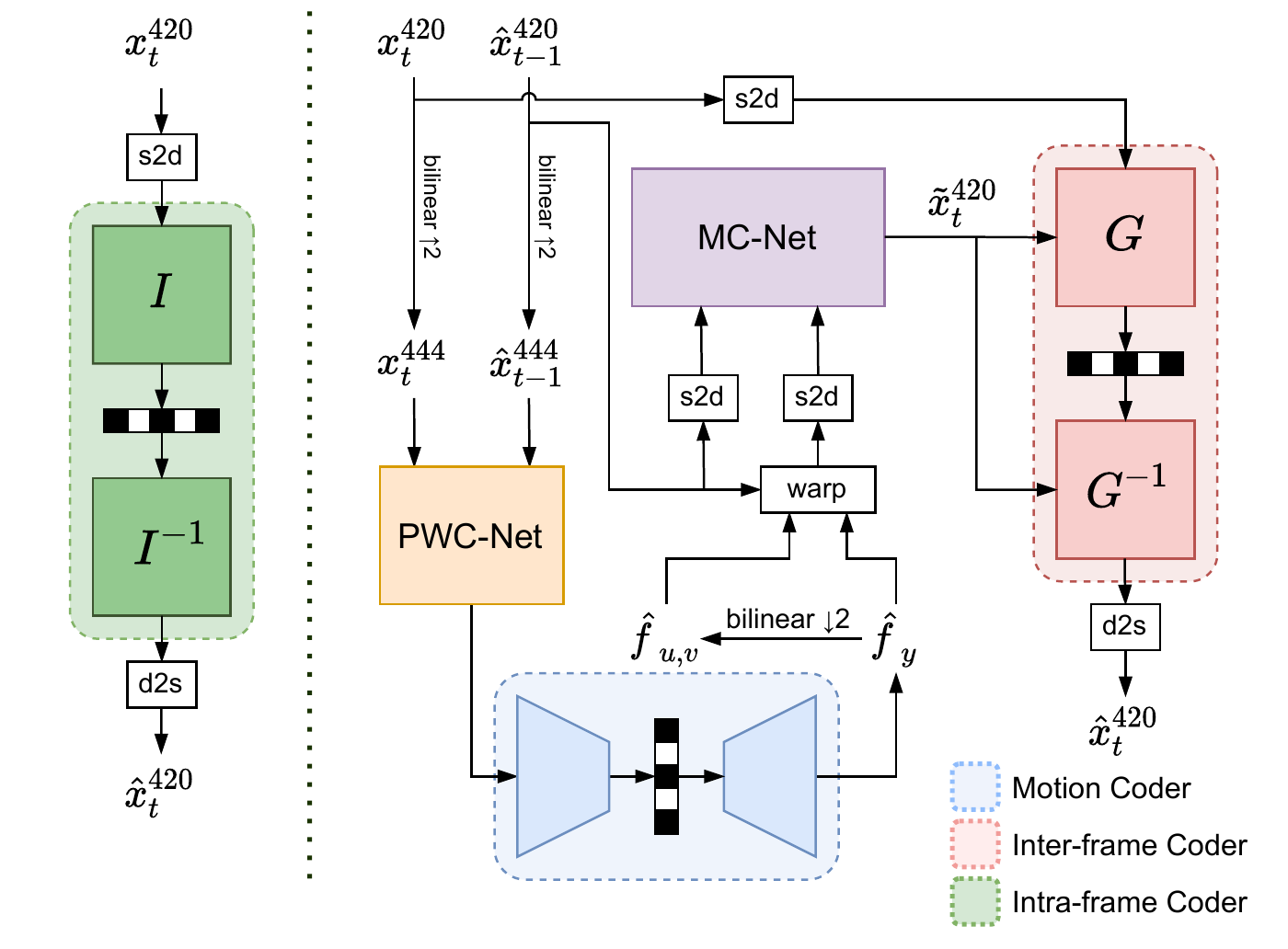}
\caption{The overall architecture of our proposed framework. The left-hand side shows the I-frame coder, and the right-hand side shows the P-frame coder. $x^{420}_t$ denotes the current coding frame and $\hat{x}^{420}_{t-1}$ means the previously reconstructed frame. To deal with YUV 4:2:0 format, several space-to-depth (s2d) and depth-to-space (d2s) operations are performed. \textcolor{black}{It is worth noting that $\tilde{x}^{420}_t$ is a 6-channel format.}}
\label{fig:arch}
\vspace{-0.5cm}
\end{figure}

Fig.~\ref{fig:arch} depicts our proposed system for coding YUV 4:2:0 content. As shown, it comprises an I-frame coder (the left part of Fig.~\ref{fig:arch}) and a P-frame coder (the right part of Fig.~\ref{fig:arch}). We adopt ANFIC~\cite{anfic} as our I-frame coder.
However, ANIFC is design primarily for RGB content; it needs to be adapted to YUV 4:2:0 content. To this end, we apply the space-to-depth (s2d) operation to the Y component, in order to convert it into a 4-channel signal that has the same spatial resolution as the UV components. The resulting signal is then concatenated with the UV components to form a 6-channel input. Whenever appropriate, we perform the depth-to-space (d2s) operation to recover the Y component in its original spatial resolution. 



Our P-frame coder consists of the motion module and the inter-frame coder ($G,G^{-1}$ in Fig.\ref{fig:arch}). The motion module includes three networks: the motion estimation network (PWC-Net), the motion coder, and the motion compensation network (MC-Net). These networks serve to synthesize the prediction frame $\tilde x_t$. The process begins with PWC-Net~\cite{pwc} estimating a dense optical flow map between $x^{420}_t$ and $\hat{x}^{420}_{t-1}$. In particular, PWC-Net performs flow estimation in YUV 4:4:4 domain, where the UV components are first up-sampled and concatenated with the Y component as input ($x_t^{444}$ and $\hat{x}_{t-1}^{444}$ in Fig.~\ref{fig:arch}). This is because we want to minimize the effort to fine-tune PWC-Net, which is initially designed for 4:4:4 content. 

For the flow map coding, we adopt a motion coder similar to that of DVC\_Pro~\cite{dvcpro}. In particular, the warping of the UV components takes $\hat{f}_{uv}$, which is downsampled bilinearly from the decoded motion $\hat{f}_y$. Both the warped frame and the previously decoded frame undergo the space-to-depth (s2d) operation before they are fed to the MC-Net to generate the 6-channel motion-compensated frame $\tilde {x}_t^{420}$.


The purpose of the inter-frame coder is to encode $x^{420}_t$ conditionally based on the motion-compensated frame $\tilde{x}^{420}_t$, without evaluating explicitly a residual frame. We modify ANFIC~\cite{anfic}, which is designed for learning the unconditional distribution of images, to learn the conditional distribution $p(x^{420}_t|\tilde{x}^{420}_t)$ in two ways. First, the encoding transforms of ANFIC are conditioned on $\tilde{x}^{420}_t$ by concatenating inputs of every encoding transforms with $\tilde{x}^{420}_t$. Second, instead of requiring $p(x_2)$ to follow the standard Normal $\mathcal{N}(0,I)$ as for learning an unconditional distribution with ANFIC~\cite{anfic}, we now require $p(x_2)$ be governed by $\mathcal{N}(\tilde{x}^{420}_t,I)$. In other words, the encoding process of our inter-frame coder is to transform the input $x^{420}_t$ into a single approximating $\tilde{x}^{420}_t$. The latent code captures the information needed to signal such transformation. 



\begin{figure}[h!]
\vspace{.6em}
\centering
\includegraphics[width=0.6\linewidth]{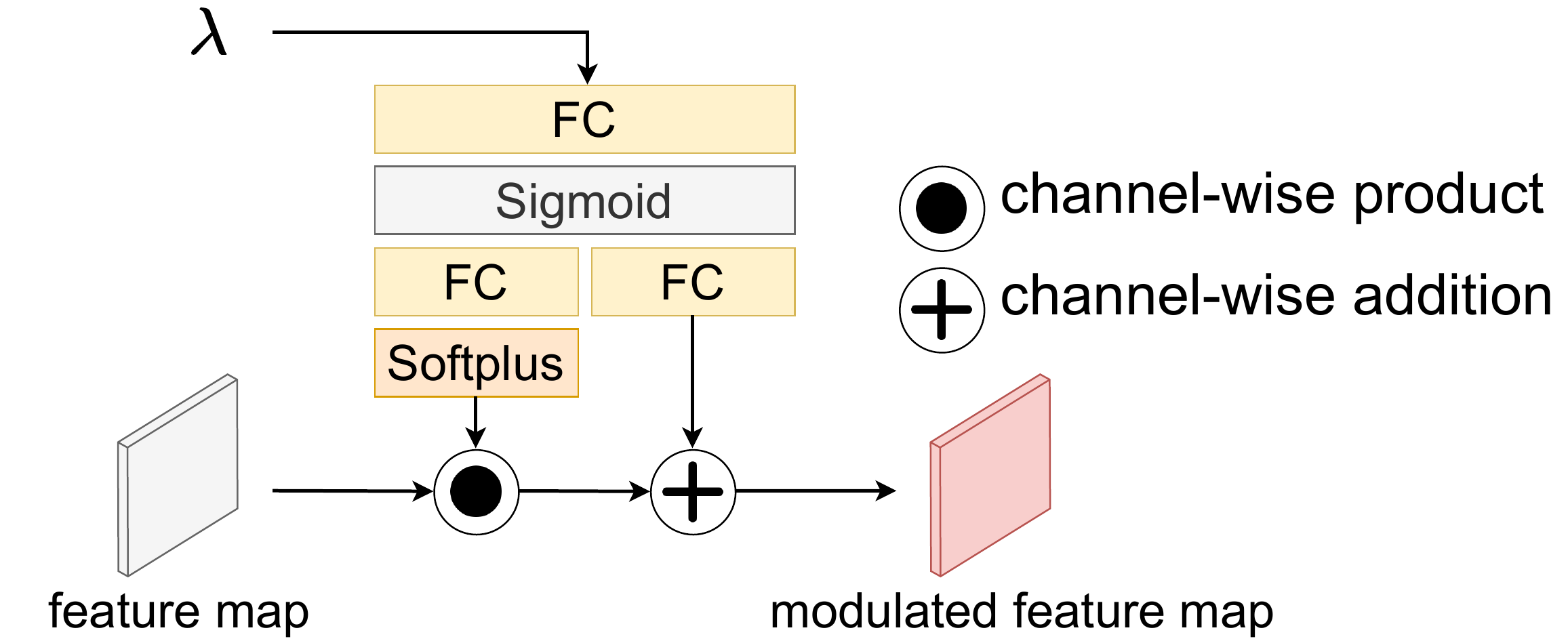}
\caption{The architecture of the rate-adaption net. The rate-adaption net takes $\lambda$ as input condition, which can be an one-hot vector or a scalar variable. The output feature can be modulated by conditional input.}
\label{fig:rate-ad-net}
\vspace{-0.5cm}
\end{figure}

\begin{table}[]
\vspace{1em}
\centering
\caption{$\lambda_I$ and $\lambda_P$ for variable-rate encoding.} 
\label{tab:lambda_range}
\resizebox{\linewidth}{!}{
\begin{tabular}{|c|c|c|c|c|}
\hline
$\lambda_P$ & 1024                   & 4096                   & 16384                  & 65536                  \\ \hline
$\lambda_I$ & $5e^{-3} \sim 5e^{-2}$ & $1e^{-2} \sim 1e^{-1}$ & $2e^{-2} \sim 2e^{-1}$ & $2e^{-1} \sim 5e^{-1}$ \\ \hline
\end{tabular}
}
\vspace{-2em}
\end{table}

\subsection{Variable-rate Encoding}
\label{Sec:VariableRate}
To achieve variable-rate encoding with a single network, we use a rate-adaption net to modulate the latent features. Specifically, the rate-adaption net outputs a set of affine transformation parameters, which are applied channel-wisely to feature maps in every convolutional layers. It adapts the dynamic ranges of feature maps according to the $\lambda$ value for the rate-distortion trade-off. Fig. \ref{fig:rate-ad-net} shows the detailed architecture of the rate-adaption net, where the input $\lambda$ can be a one-hot vector for discrete-step rate adaption or a real-valued scalar for continuous-step rate adaption. In our implementation, we adopt continuous-step rate adaption on the I-frame coder and discrete-step rate adaption on the P-frame coder. In particular, the $\lambda$ values of the I-frame coder $\lambda_I$ are partitioned into 4 overlapping groups, with each group corresponding to a particular $\lambda_P$ value for the P-frame coder (see Table \ref{tab:lambda_range}). At inference time, to encode a video at a target bit rate, we first choose the combination of ($\lambda_I$,$\lambda_P$) which yields a rate point matching closely the target. We then fine-tune $\lambda_I$ continuously while fixing the chosen $\lambda_P$ until the target rate is met.




\begin{figure*}[h]
\begin{center}
\begin{subfigure}{0.39\linewidth}
    \centering
    \hspace{-2em}
    \includegraphics[width=0.9\linewidth]{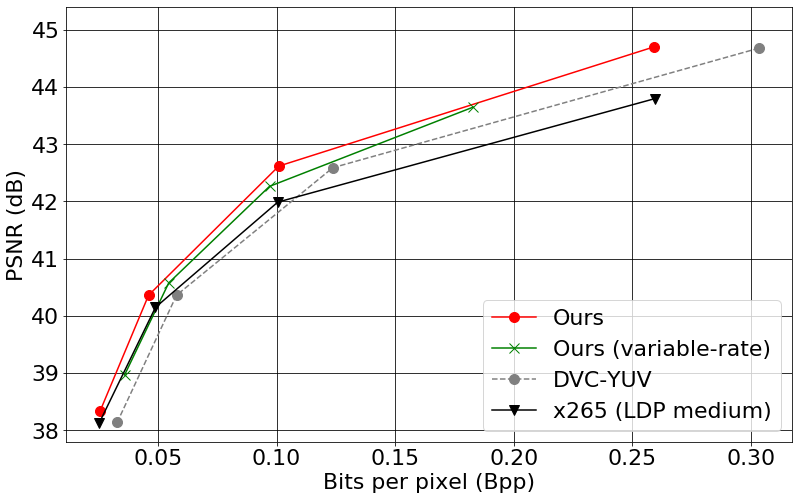}
    \vspace{-0.6em}
    \caption{UVG, PSNR}
    \label{fig:uvgPSNR}
\end{subfigure}
\begin{subfigure}{0.39\linewidth}
    \centering
    \includegraphics[width=0.9\linewidth]{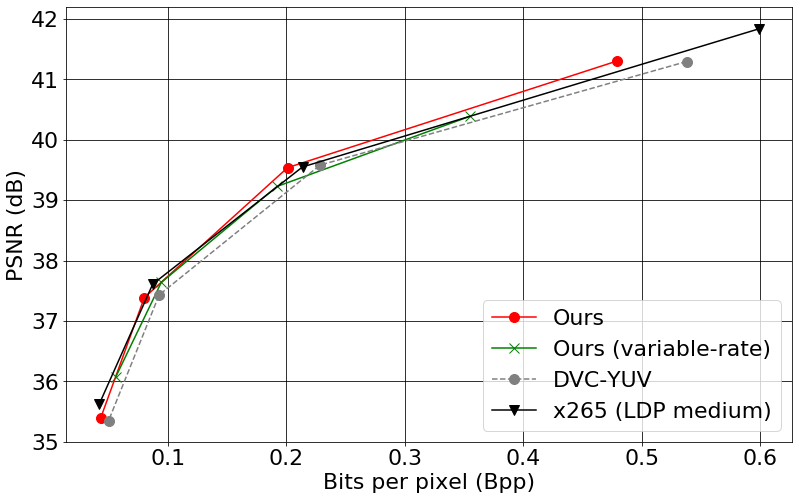}
    \vspace{-0.6em}
    \caption{HEVC Class B, PSNR}
    \label{fig:hevcPSNR}
\end{subfigure}

\begin{subfigure}{0.39\linewidth}
    \centering
    \hspace{-2em}
    \includegraphics[width=0.9\linewidth]{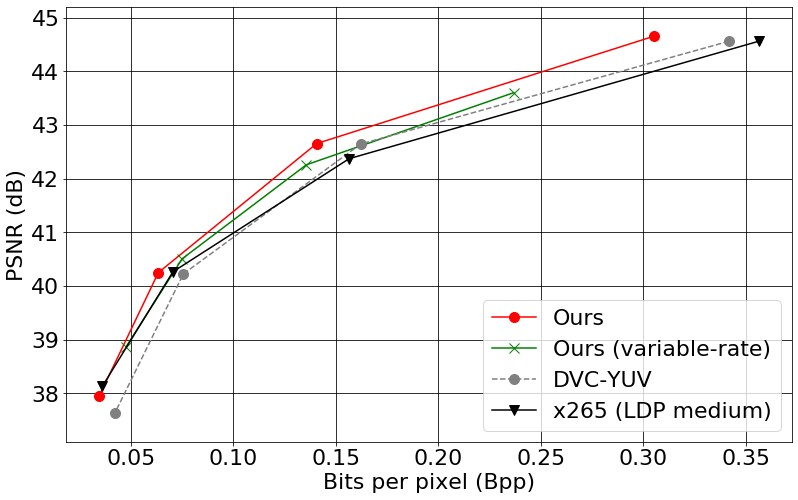}
    \vspace{-0.6em}
    \caption{MCL-JCV, PSNR}
    \label{fig:mclPSNR}
\end{subfigure}
\begin{subfigure}{0.39\linewidth}
    \centering
    \includegraphics[width=0.9\linewidth]{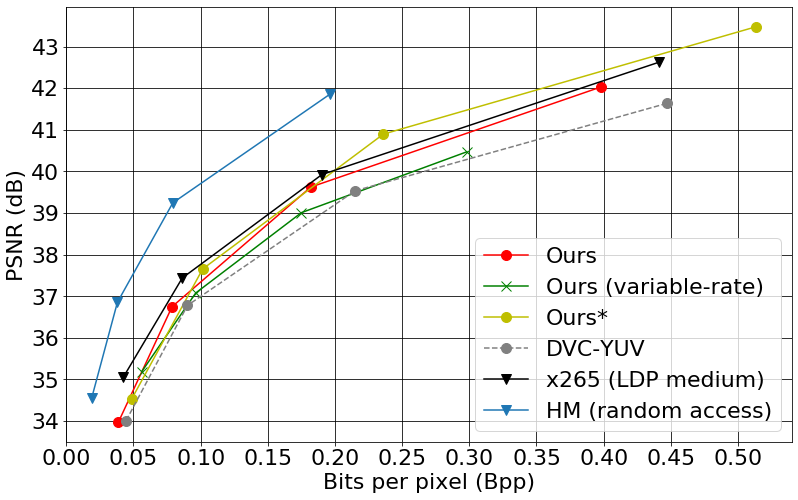}
    \vspace{-0.6em}
    \caption{ISCAS'22 GC, PSNR}
    \label{fig:gcPSNR}
\end{subfigure}
\caption{Rate-distortion performance evaluation on UVG, HEVC Class B, MCL-JCV, and ISCAS'22 GC datasets.}
\label{fig:RD}
\end{center}
\vspace{-2.0em}
\end{figure*}
\subsection{Training Procedure}\label{AA}
{\bf I-frame Coder:}
To train the I-frame coder, we follow the training procedure in~\cite{anfic} and train the model for the highest rate point with $\lambda_I = 5e^{-1}$. We then enable the rate-adaption net to train the variable-rate model by minimizing:
\vspace{0.5em}
\begingroup
\setlength\abovedisplayskip{0pt}
\begin{equation*}
\label{equ:rd_loss_I}
 L = \frac{1}{N}\sum_{\lambda_I\sim\Lambda_I}{R+ \lambda_I \cdot(2 MSE_Y+MSE_U+MSE_V)/4},
\vspace{-0.3em}
\end{equation*}
\endgroup

where $\lambda_I$ is chosen randomly from the set $\Lambda_I=[5e^{-3}, 5e^{-1}]$, and $N$ denotes the batch size. 


{\bf P-frame Coder:} 
In a similar way to training the I-frame coder, we first train the single-rate P-frame coder and then fine-tune it for the variable-rate case. For the single-rate case, the training is done sequentially: the motion module is updated first, followed by updating the conditional inter-frame coder while fixing the motion module. Next, the whole system is fine-tuned end-to-end and jointly. 

For training the variable-rate model, we take the single-rate model as the pre-trained model. We then update the rate-adaption net by enabling it in all the convolutional layers inside the conditional inter-frame coder, while fixing the pre-trained weights. In the next training stage, the rate-adaption net is applied to the motion module, where the motion estimation network is kept untouched. Again, only the rate-adaption net is updated in this stage. Finally, the whole system is fine-tuned end-to-end. 
The training objective of the P-frame coder is:
\vspace{0.5em}
\begingroup
\setlength\abovedisplayskip{0pt}
\begin{equation*}
\label{equ:rd_loss_P}
 L = \frac{1}{N}\sum_{\lambda_P\sim\Lambda_P}{R + \lambda_P \cdot (2 MSE_Y+MSE_U+MSE_V)/4},
\vspace{-0.3em}
\end{equation*}
\endgroup
where $\lambda_P$ is chosen randomly from $\Lambda_P=\{1024, 4096, 16384, 65536\}$ in a mini-batch. It is worth nothing that when training the P-frame coder, the reconstructed I-frame is regarded as a constant; no gradient will be back-propagated to the I-frame coder.








\section{Experiments}

{\bf Implementation Details:} The architecture of our intra-frame coder is similar to ANFIC ~\cite{anfic} (see N, M, K, L in Fig.~\ref{fig:anfic}), but with only three Conv and two GND layers due to s2d operation. The inter-frame coder has a similar model architecture to the intra-frame coder, but has additional conditioning variables concatenated to the input of each encoding transform and different K and L (see Fig.~\ref{fig:anfic}). The motion coder has exactly the same architecture with hyperprior~\cite{googleiclr18} with N = M = k = L = 128. For training, we use Vimeo-90k dataset~\cite{vimeo}. Since Vimeo-90k is in RBG format, we generate the training data by converting Vimeo-90k into YUV 4:2:0 format with resolution 448x256. During training, we randomly crop the frames to 256x256, so the sizes of the chroma components are 128x128. We adopt Adam optimizer~\cite{adam}, and the learning rate is fixed at $1e^{-4}$ before 300k iterations, and is decreased to $1e^{-5}$ then.


{\bf Evaluation Methodologies:} 
For evaluation, we test our scheme on UVG~\cite{uvg}, MCL-JVC~\cite{mcl}, HEVC Class B~\cite{hevcctc} and the test dataset provided by the Grand Challenge (GC)~\cite{iscas_gc}. All the test sequences are in YUV 4:2:0 format. We follow the common test protocol to set GOP size to 12 for UVG~\cite{uvg} and MCL-JVC~\cite{mcl}, 10 for HEVC Class B~\cite{hevcctc}, and 32 for ISCAS'22 Grand Challenge (GC)~\cite{iscas_gc}. The PSNR is measured according to $PSNR = (6PSNR_Y + PSNR_U + PSNR_V)/8$ and the bit-rate is measured in bits per pixel (bpp).

{\bf Baseline methods:}
To generate the \textit{x265} baseline results, we use \textit{ffmpeg}~\cite{ffmpeg} with \textit{medium} preset and \textit{low delay} configuration. The \textit{QPs} are set to 22, 27, 32 and 37. 
For the learning-based baseline, we train DVC-YUV, which has the same intra-frame and motion coders as our scheme but replaces the ANF-based inter-frame conditional coder with the VAE-based residual coder~\cite{dvcpro}. To make a fair comparison, we expand the channel number of the residual coder of DVC-YUV to N = 192, so that DVC-YUV and our proposed model have comparable model sizes.

{\bf Experimental Results:}
Fig.~\ref{fig:RD} and Table~\ref{tab:exp_BD_PSNR} show the rate-distortion performance of our proposed single-rate model, variable-rate model, x265, and DVC-YUV. 

On UVG (Fig.~\ref{fig:uvgPSNR}) and MCL-JCV datasets (Fig. \ref{fig:mclPSNR}), both our single-rate model and variable-rate model show better rate-distortion performance than x265. A significant improvement is observed at higher rates, while comparable performance can be seen at lower rates. In terms of BD-rate savings (Table~\ref{tab:exp_BD_PSNR}), the single-rate model achieves 18\% and 13.1\% rate reductions; in contrast, the multi-rate model shows 10.9\% and 4.5\% rate reductions.


On HEVC Class B dataset (Fig.~\ref{fig:hevcPSNR}), our models show comparable performance to x265 at both high rates and low rates, resulting in 1\% overall rate reductions with the single-rate model and 7.7\% rate inflation with the variable-rate model. 
In particular, our model shows 9.6\% rate inflation on the ISCAS'22 GC test dataset as compared with x265 (see Fig.~\ref{fig:gcPSNR}). It is worth noting that on this dataset, a much larger GOP size of 32 is used, as compared to 10 or 12 on UVG, MCL-JCV, and HEVC Class B datasets. The worse performance of our model is due to the use of the less capable inter-frame coder and the training strategy. To see this, we additionally train a more powerful model (denoted as Ours*) with the channel numbers N = M = k = L = 192. We also include the temporal prior~\cite{dcvc} and follow~\cite{epr} to train several additional epochs to alleviate error propagation for large GOP's. As can be seen from Fig.~\ref{fig:gcPSNR} and Table~\ref{tab:exp_BD_PSNR}, the enhanced model (Ours*) achieves better performance than x265 at higher rates. However, at lower rates where the motion overhead plays a more critical role, there is still room for improvement. Nevertheless, our models outperform DVC-YUV by a significant margin on all the datasets. Note that there is still a large gap between the HM Random Access anchor and our scheme. Apparently, bi-prediction is a tool that needs to be incorporated.

\begin{table}[]
    \centering
    \caption{BD-rate comparison with x265 (LDP medium) serving as the anchor.}
    \label{tab:exp_BD_PSNR}
    \centerline{%
    \resizebox{1\linewidth}{!}{
    \begin{tabular}{lcccc}
        \toprule
        \multirow{2}{*}{Methods}  & 
            \multicolumn{4}{c}{BD-rate (\%) PSNR} \\
            \cline{2-5} 
                                & UVG    & HEVC-B & MCL-JCV & ISCAS'22 GC \\ 
                \hline
        \textbf{Ours} & -18.0 & -1.0 & -13.1 & 9.6  \\
        \textbf{Ours (variable-rate)}& -10.9 & 7.7 & -4.5 &	24.9 \\
        \textbf{DVC-YUV}& 3.2 & 12.3 & 3.6 &  29.6 \\
        \textbf{Ours*}& -	& - & - &	3.1 \\
        \bottomrule
        \vspace{0.1em}
    \end{tabular}
    }%
    }
\vspace{-3.6em}
\end{table}



\section*{Conclusion}
In this paper, we propose a learning-based conditional inter-frame coding scheme for YUV 4:2:0 video. Our experimental results show that the proposed scheme can outperform x265 on UVG and MCL-JCV. However, on the more challenging datasets from ISCAS'22 GC, there is still ample room for improvement. One reason of the inferior performance of our model on this dataset is insufficient infer-frame coding at a large GOP size, which can be improved by increasing the model capacity and applying an error propagation-aware training strategy. In addition, how to enhance the motion coder to improve the low rate performance is among our future work.




 
{\small
\bibliographystyle{ieee_fullname}
\bibliography{egbib}

\begin{thebibliography}{10}\itemsep=-1pt

\bibitem{ffmpeg}
"ffmpeg software".
\newblock {\em URL http://ffmpeg.org/}.

\bibitem{iscas_gc}
"grand challenge on neural network-based video coding".
\newblock {\em URL https://www.iscas2022.org/grand-challenge}, 2021.

\bibitem{googleiclr18}
Johannes Ball{\'e}, David Minnen, Saurabh Singh, Sung~Jin Hwang, and Nick
  Johnston.
\newblock Variational image compression with a scale hyperprior.
\newblock In {\em International Conference on Learning Representations}, 2018.

\bibitem{hevcctc}
Frank Bossen et~al.
\newblock Common test conditions and software reference configurations.
\newblock {\em JCTVC-L1100}, 12(7), 2013.

\bibitem{adam}
Jimmy~Ba Diederik P.~Kingma.
\newblock Adam: A method for stochastic optimization.
\newblock {\em International Conference for Learning Representations}, 2015.

\bibitem{anfic}
Yung-Han Ho, Chih-Chun Chan, Wen-Hsiao Peng, Hsueh-Ming Hang, and Marek
  Domański.
\newblock Anfic: Image compression using augmented normalizing flows.
\newblock {\em IEEE Open Journal of Circuits and Systems}, 2:613--626, 2021.

\bibitem{fvc}
Zhihao Hu, Guo Lu, and Dong Xu.
\newblock Fvc: A new framework towards deep video compression in feature space.
\newblock In {\em Proceedings of the IEEE/CVF Conference on Computer Vision and
  Pattern Recognition}, pages 1502--1511, 2021.

\bibitem{mmsp}
Th{\'e}o Ladune, Pierrick Philippe, Wassim Hamidouche, Lu Zhang, and Olivier
  D{\'e}forges.
\newblock Optical flow and mode selection for learning-based video coding.
\newblock In {\em 2020 IEEE 22nd International Workshop on Multimedia Signal
  Processing (MMSP)}, pages 1--6. IEEE, 2020.

\bibitem{dcvc}
Jiahao Li, Bin Li, and Yan Lu.
\newblock Deep contextual video compression.
\newblock {\em Advances in Neural Information Processing Systems}, 34, 2021.

\bibitem{mlvc}
Jianping Lin, Dong Liu, Houqiang Li, and Feng Wu.
\newblock M-lvc: multiple frames prediction for learned video compression.
\newblock In {\em Proceedings of the IEEE/CVF Conference on Computer Vision and
  Pattern Recognition}, pages 3546--3554, 2020.

\bibitem{icip21}
Jianping Lin, Dong Liu, Jie Liang, Houqiang Li, and Feng Wu.
\newblock A deeply modulated scheme for variable-rate video compression.
\newblock In {\em IEEE International Conference on Image Processing}, 2021.

\bibitem{epr}
Guo Lu, Chunlei Cai, Xiaoyun Zhang, Li Chen, Wanli Ouyang, Dong Xu, and Zhiyong
  Gao.
\newblock Content adaptive and error propagation aware deep video compression.
\newblock In {\em European Conference on Computer Vision}, pages 456--472.
  Springer, 2020.

\bibitem{dvclu}
Guo Lu, Wanli Ouyang, Dong Xu, Xiaoyun Zhang, Chunlei Cai, and Zhiyong Gao.
\newblock Dvc: An end-to-end deep video compression framework.
\newblock In {\em Proceedings of the IEEE/CVF Conference on Computer Vision and
  Pattern Recognition}, pages 11006--11015, 2019.

\bibitem{dvcpro}
Guo Lu, Xiaoyun Zhang, Wanli Ouyang, Li Chen, Zhiyong Gao, and Dong Xu.
\newblock An end-to-end learning framework for video compression.
\newblock {\em IEEE transactions on Pattern Analysis and Machine Intelligence},
  2020.

\bibitem{uvg}
Alexandre Mercat, Marko Viitanen, and Jarno Vanne.
\newblock Uvg dataset: 50/120fps 4k sequences for video codec analysis and
  development.
\newblock In {\em Proceedings of the 11th ACM Multimedia Systems Conference},
  pages 297--302, 2020.

\bibitem{elfvc}
Oren Rippel, Alexander~G. Anderson, Kedar Tatwawadi, Sanjay Nair, Craig Lytle,
  and Lubomir Bourdev.
\newblock Elf-vc: Efficient learned flexible-rate video coding.
\newblock In {\em Proceedings of the IEEE/CVF International Conference on
  Computer Vision (ICCV)}, pages 14479--14488, October 2021.

\bibitem{pwc}
Deqing Sun, Xiaodong Yang, Ming-Yu Liu, and Jan Kautz.
\newblock Pwc-net: Cnns for optical flow using pyramid, warping, and cost
  volume.
\newblock In {\em Proceedings of the IEEE conference on computer vision and
  pattern recognition}, pages 8934--8943, 2018.

\bibitem{mcl}
Haiqiang Wang, Weihao Gan, Sudeng Hu, Joe~Yuchieh Lin, Lina Jin, Longguang
  Song, Ping Wang, Ioannis Katsavounidis, Anne Aaron, and C-C~Jay Kuo.
\newblock Mcl-jcv: a jnd-based h. 264/avc video quality assessment dataset.
\newblock In {\em 2016 IEEE International Conference on Image Processing
  (ICIP)}, pages 1509--1513. IEEE, 2016.

\bibitem{vimeo}
Tianfan Xue, Baian Chen, Jiajun Wu, Donglai Wei, and William~T Freeman.
\newblock Video enhancement with task-oriented flow.
\newblock {\em International Journal of Computer Vision}, 127(8):1106--1125,
  2019.

\bibitem{hlvc}
Ren Yang, Fabian Mentzer, Luc~Van Gool, and Radu Timofte.
\newblock Learning for video compression with hierarchical quality and
  recurrent enhancement.
\newblock In {\em Proceedings of the IEEE/CVF Conference on Computer Vision and
  Pattern Recognition}, pages 6628--6637, 2020.

\bibitem{cconv}
Jungwon~Lee Yoojin~Choi, Mostafa El-Khamy.
\newblock Variable rate deep image compression with a conditional autoencoder.
\newblock In {\em International Comference on Computer Visions}, 2019.

\end{thebibliography}
}

\end{document}